\documentclass[aps,prl,twocolumn,superscriptaddress]{revtex4-1}

\usepackage{graphicx}
\usepackage{longtable}

\begin{document}


\title{Gapless excitations in the ground state of 1T-TaS$_2$}

\author{A. Ribak}
\affiliation{Physics Department, Technion-Israel Institute of Technology, Haifa 32000, Israel}
\author{I. Silber}
\affiliation{Raymond and Beverly Sackler School of Physics and Astronomy, Tel-Aviv University, Tel Aviv, 69978, Israel}
\author{C. Baines}
\affiliation{Paul Scherrer Institute, CH 5232 Villigen PSI, Switzerland}
\author{K. Chashka}
\affiliation{Physics Department, Technion-Israel Institute of Technology, Haifa 32000, Israel}
\author{Z. Salman}
\affiliation{Paul Scherrer Institute, CH 5232 Villigen PSI, Switzerland}
\author{Y. Dagan}
\affiliation{Raymond and Beverly Sackler School of Physics and Astronomy, Tel-Aviv University, Tel Aviv, 69978, Israel}
\author{A. Kanigel}
\affiliation{Physics Department, Technion-Israel Institute of Technology, Haifa 32000, Israel}




\date{\today}

\begin{abstract}
1T-TaS$_2$ is a layered transition metal dichalgeonide with a very rich phase diagram. At T=180K it undergoes a metal to Mott insulator transition. Mott insulators usually display anti-ferromagnetic ordering in the insulating phase but 1T-TaS$_2$ was never shown to order magnetically. In this letter
we show that 1T-TaS$_2$ has a large paramagnetic contribution to the magnetic susceptibility but it does not show any sign of magnetic ordering or freezing down to 20mK, as probed by $\mu$SR, possibly indicating a  quantum spin liquid ground state. Although 1T-TaS$_2$ exhibits a strong resistive behavior both in and out-of plane at low temperatures we find a linear term in the heat capacity suggesting the existence of a Fermi-surface, which has an anomalously strong magnetic field dependence.
\end{abstract}


\maketitle

\section{}
\preprint{twocolumn}
The elusive quantum spin liquid (QSL) state has been the subject of numerous papers since Anderson suggested the resonating valence bond (RVB) as the ground state of the $S=1/2$ quantum Heisenberg antiferromagnet on a triangular lattice \cite{Anderson_RVB}.
A QSL is an exotic state of matter with an insulating ground state that does not break the crystal symmetry. The spins are highly entangled, and quantum fluctuations prevent magnetic ordering down to absolute zero T=0K. Theoretically, it has been suggested that a QSL can support exotic spinon excitations \cite{Anderson_RVB,Anderson_Fazekas,Balents_QSL_review}.


The experimental search for QSL materials has been focused on $S=1/2$ quantum spins on frustrated lattices with triangular motifs. Despite extensive research, very few candidates exist: the 2D kagom\'{e} hebertsmithite and vesignieite \cite{Hebertsmithite,QSLexp_SUS3}, the 2D triangular lattices $\kappa$-(BEDT-TTF)$_2$Cu$_2$(CN)$_3$,
EtMe$_3$Sb[Pd(dmit)$_2$]$_2$ and YbMgGaO$_4$  \cite{KBEDT_PRL_2003,EtME3_2007,KBDET_NatPhys_2008,QSL_exp_SUS1,KBEDT_2011,YbMgGaO_2016,YbMgGaO_2017}, and the 3D hyperkagom\'{e} Na$_4$Ir$_3$O$_8$  \cite{Na4Ir3O8_PRL2007,Na4Ir3O8_2008}.

Recently, 1T-TaS$_2$ has been suggested to have a QSL ground state \cite{Lee}. 1T-TaS$_2$ has been a major subject of interest for over 40 years owing to its very rich phase diagram, arising from  strong electron-electron and electron-phonon couplings \cite{DiSalvo_long,Rossnagel2011}. At high temperatures ($>550 $K) the system is metallic, and it undergoes a series of charge density wave (CDW) transitions as the temperature is lowered. It can even become superconducting when subjected to external pressure or chemical disorder \cite{Sipos,Gironi_Pres_SC}.

Despite the seemingly complicated electronic properties, 1T-TaS$_2$ has a simple crystal structure composed of weakly bound van der Waals layers; each layer contains a single sheet of Tantalum (Ta) atoms, sandwiched in between two sheets of Sulfur (S) atoms in an octahedral coordination. The Ta atoms within each layer form a 2D hexagonal lattice.

\begin{figure}[h!]
	\centering	
	\includegraphics[width=86mm]{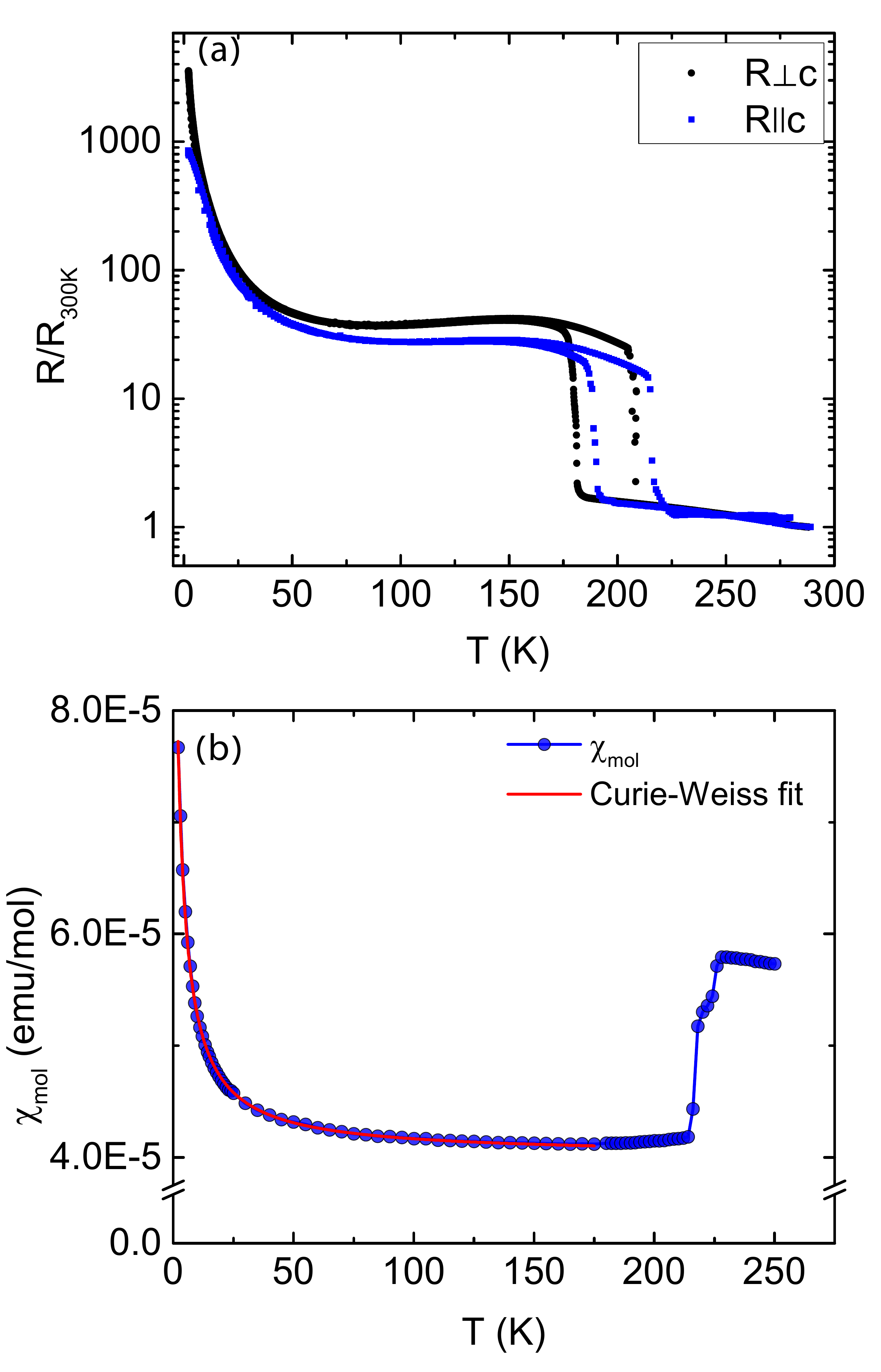}
	\caption{(a) Normalized resistance versus temperature. The black (blue) curve is in-plane (out-of-plane) normalized resistance, all measured in 4-contacts configuration. (b) Magnetic molar susceptibility versus temperature, obtained from SQUID magnetization measurements at constant magnetic fields of 1 and 3 T. The red line is the fit to Curie-Weiss form at the temperature range of 2 to 175 K.}
	\label{Fig1}
\end{figure}

The basic CDW instability is formed within the Ta layers by the arrangement of 13 Ta atoms into a "Star-of-David" shaped cluster, where 12 Ta atoms move slightly inwards towards the 13th central Ta atom. In the temperature range of $\sim$350-550K the system is in the incommensurate (IC) CDW phase. When cooled below $\sim$350K, the system is in the nearly commensurate (NC) state, where the CDW clusters lock-in to form commensurate domains, and the domains in turn form a triangular lattice. The size of the domains becomes larger as temperature is decreased until all the domains interlock into a single coherent CDW modulation extending throughout the layer.
This transition into the commensurate CDW phase (CCDW) occurs at T$_{CCDW}\sim$180 K and is accompanied by a metal-insulator (MI) transition. The MI transition is easily visible in resistance measurements (as an abrupt jump in the resistance; see Fig. \ref{Fig1}a). Below T$_{CCDW}$, the large negative slope in resistance as a function of the temperature indicates a strongly insulating behavior. Heating the sample reveals an hysteretic behavior expected from a first order transition.

In the CCDW phase the unit cell is reconstructed into a rotated triangular lattice with a unit cell of size $\sqrt{13}\times\sqrt{13}$ of the original lattice. In the reconstructed lattice, every site is the center of the 13-atom "Star-of-David" cluster. Each Ta atom has a 4$+$ valence and a single $5d$ electron, thus we have a single unpaired electron per cluster. Based on band theory the ground state should be metallic, however, interactions localize the electrons, driving the system into a Mott insulating state \cite{Fazekas_Tosatti_1980}. A Mott insulator is expected to be an antiferromagnet since when neighboring spins are oppositely aligned; one can gain energy of $J=4t^2/U$ by virtual hopping. This exchange energy, $J$, is the effective interaction between two neighbor spins. So far magnetic-ordering has never been observed in 1T-TaS$_2$ \cite{Fazekas1980,DiSalvo_Sus_ESR,Furukawa,Perfetti2005}.

Evidence for a QSL state are usually circumstantial and it is usually easier to rule out other possible ground states and suggest a QSL by elimination. The necessary ingredients for a QSL state are: \textit{(A)} A charge-gap, as the material should be insulating at the ground state. \textit{(B)} The existence of an odd number of unpaired spins (electrons) per unit cell that \textit{(C)} do not order down to 0 K. 


High-quality single-crystals of 1T-TaS$_2$ were grown using standard chemical vapor transport method \cite{DiSalvo_CVT} with Iodine as transport gas. The crystal structure and chemical compositions were verified using X-ray diffraction and energy-dispersive X-ray spectroscopy (EDS). No spurious phases were found, in particular, none of the other TaS$_2$ polymorphs is present in the crystals (see supplementary material).

We present the in-plane (black) and out-of-plane (blue) normalized resistance (R) versus temperature in Fig. \ref{Fig1}a. Both curves exhibit the same behavior with a sharp increase in R at 180 K when cooled down, due to the hysteretic first order phase transition from the nearly-commensurate to commensurate phase. When the temperature is lowered R shoots up and the material is said to be in the Mott insulating state. We find that large normalized resistance, R$_{2K}/$R$_{300K}$ is indicative of high quality crystals.  This ratio is very sensitive to the crystal growth conditions. All the data we present here were measured on crystals with R$_{2K}/$R$_{300K} > 3000$ (in the in-plane direction).

The temperature dependence of the magnetization was measured using a SQUID magnetometer in magnetic fields of 1 and 3 T. The DC magnetic susceptibility versus temperature was extracted by a linear fit with zero intercept, and is presented in Fig. \ref{Fig1}b.
The data are shown after the subtraction of the core diamagnetic contribution \cite{magnetochem}. Several observations can be made from these data. 1) At T$\approx$210 K the susceptibility decreases sharply due to the CCDW transition. The decrease is a result of the gap opening and the loss of the Pauli contribution to the susceptibility due to the reduction of density of state at the Fermi level. 2) Between 50 K and 2 K the susceptibility increases monotonically, exhibiting a paramagnetic Curie-like behaviour.
3) There is a paramagnetic temperature-independent contribution to the susceptibility.

We fit a Curie-Weiss model with an additional constant to the data in the temperature range of 1.6 - 125 K (red curve in Fig. \ref{Fig1}b):
\begin{equation}
\chi=\frac{C}{T-\theta_{CW}}+\chi_0.
\label{CW_eq}
\end{equation}

The best fit results in $\theta_{CW}=$-2.1$\pm$0.2 K , $C=$1.53E-4 $\pm$1.2E-6 emu K/mol and $\chi_0$=4E-5 $\pm$ 3.7E-8 emu/mol. The Curie constant, $C$, depends on the spin volume density, $n$, and the $g$-factor:
\begin{equation}
C=\frac{ng^2S(S+1)\mu_B^2}{3k_B}=\frac{n\mu^2}{3k_B},
\label{Curie_const}
\end{equation}
thus from $C$ we can extract the number of spins and compare it to the number of Ta atoms in our sample for a given $g$-factor.  In order to measure the $g$-factor we have performed electron-spin resonance (ESR) at 9 GHz. In agreement with Ref. \cite{DiSalvo_Sus_ESR}, we could not find a resonance up to a magnetic field of 1.6T and down to a temperature of 5K. This suggests that either the $g$-factor is smaller than 0.44 or that the resonance is too broad to be observed.

Very recently, an effective moment $\mu_{eff}=0.4\mu_B$ was reported based on inelastic Neutrons experiments \cite{TaS2_npj}. Using this value for the effective moment in Eq. \ref{Curie_const} we find a spin concentration of one spin per 130 Ta atoms or one spin per 10 stars-of-David. Using a more conservative value of g=2.0 we get a spin concentration of 1 spin per 2450 Ta atoms or 1 spin per 188 stars-of-David. We find a spin concentration of 10 to $0.5$ percent of the expected spin density based on the Mott insulator model \cite{Fazekas_Tosatti_1980}. This concentration is  too high to be the result of spurious magnetic impurities. Moreover, our value for the Curie constant is in agreement with susceptibility measurements done by other groups \cite{DiSalvo_Sus_ESR,TaS2_sus_RHFriend,TaS2_npj,arcon}, suggesting that the Curie-like contribution is intrinsic to the 1T-TaS$_2$ crystal.

 The Cuire-Weiss fit to the data yields a Curie temperature from which the exchange coupling $J$ can be estimated using $J=\frac{3k_B\theta_{CW}}{2zS(S+1)}$ where $z$ is the number of nearest neighbours. We find $J\approx0.1$ meV, a rather small value compared to other QSL candidates \cite{Balents_QSL_review}. This $J$  represents a small  interaction between the spins that contribute to the Curie-like term and is probably much smaller than the exchange interaction between two spins residing on neighbor stars-of-David.


 The large majority of the spins (between 90\% and 99.5\% depending on the assumed g-factor) do not contribute to the Cuire-like part of the magnetic susceptibility.
One possible reason for that is that these spins are not localized. It was suggested by Darancet \textit{et al.} \cite{Millis_PRB} that there should be a substantial electronic dispersion along the c-axis and one should expect metallic conduction along that axis in the CCDW state of 1T-TaS2.

A second, more interesting, possibility is that the exchange coupling between spins in the-plane is very large, as suggested  by Law and Lee [29]. In that
case the high-temperature regime, where these spins should contribute a Curie-like term is not accessible experimentally, since $J$ is larger than k$_B$T$_{CCDW}$. In the Mott insulator phase these spins may form an anti-ferromagnetic  or  a QSL state which is responsible for
the paramagnetic temperature-independent part of the susceptibility.  Without a detailed model of the QSL it will be very difficult to  extract any valuable information from the value of $\chi_0$. In addition, the value of $\chi_0$ depends on the calculation of the core diamagnetism which is not very accurate \cite{magnetochem}.

We turn now to $\mu$SR measurements and show that 1T-TaS$_2$ shows no sign of magnetic freezing even at temperatures as low as 20 mK. $\mu$SR is a very sensitive tool which can detect small local magnetic fields originating either from long or short range order, as well as spatial inhomogeneities. The temperature dependence of the muons' polarization damping rate $(\lambda)$ is a useful measure for detecting magnetic ordering or spin freezing. Frozen local moments will result in an increase in $\lambda$.

\begin{figure}
	\centering	
	\includegraphics[width=80mm]{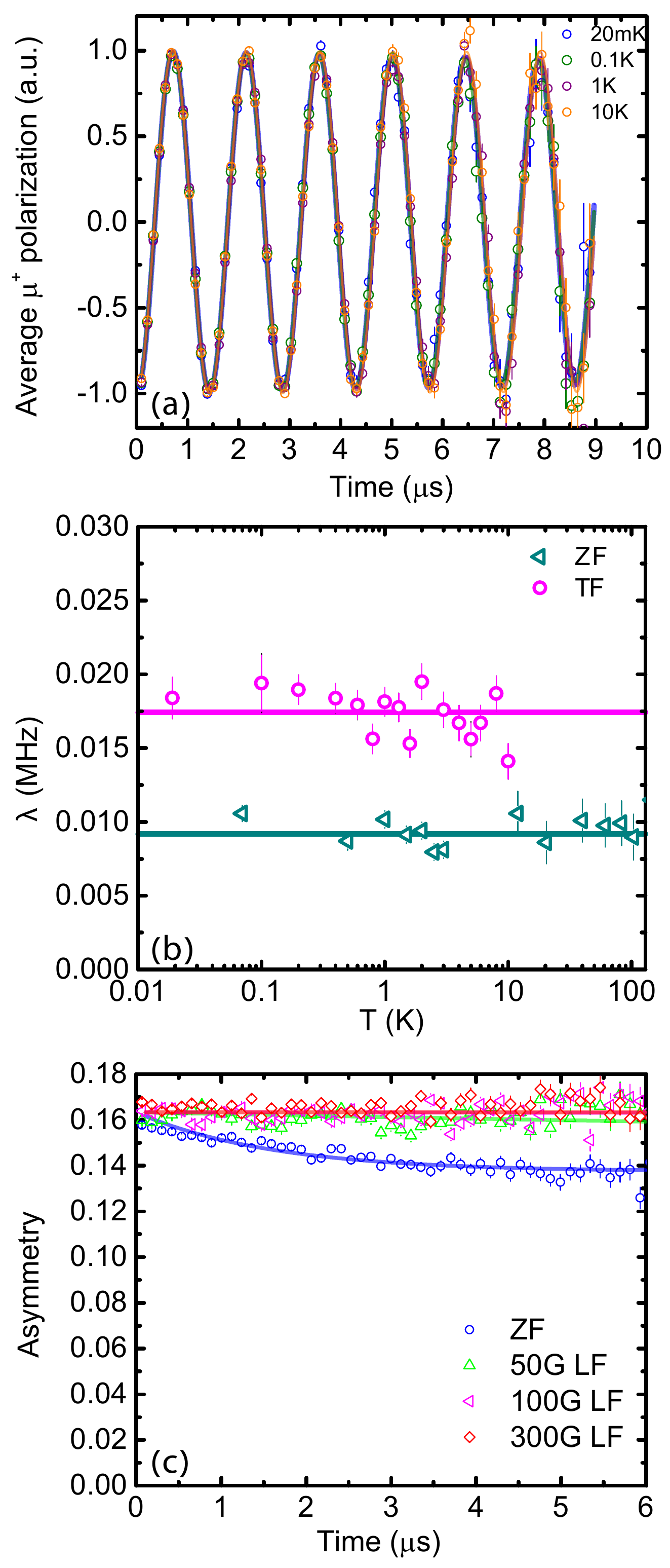}
	\caption{$\mu$SR results. (a) Averaged muon polarization at 50 G transverse field at several temperatures, below and above $\theta_{CW}$. The circles are measurement results, while the solid lines the fit results. (b) Temperature dependence of the muon spin damping rate, $\lambda$. $\lambda$ remains constant down to 20mK (TF measurements) and 70 mK (ZF measurements), implying the absence of frozen moments. The ZF asymmetry was fitted using  $P(t)=Aexp(-\lambda t)$ and the TF asymmetry was fitted using $P(t)=Aexp(-\lambda t)cos(\gamma Bt+\phi)$. The solid lines are the average $\lambda$ from the ZF and TF measurements respectively. (c) Longitudinal field $\mu$SR asymmetry, measured at 10K and at three different magnetic fields. The symbols are measurement results while the solid lines are the fit results using $P(t)=Aexp(-\lambda t)$. There is a complete decoupling of the $\mu^+$ spins from nuclear dipolar fields already at 50 G, suggesting all static moments are much smaller than 50 G.  }
	\label{muSR}
\end{figure}

We have performed zero-field (ZF) measurements at the EMU beam-line at ISIS and weak transverse-field (TF) measurements at the LTF beam-line at PSI over a wide temperature range from 200 K down to 20 mK. The TF measurements were done using a 50 Gauss field. We present the muon polarization measured in a weak TF for several temperatures in Fig. \ref{muSR}a . As can be seen there is no change in the damping.


The temperature dependence of $\lambda$ for ZF and TF is presented in Fig. \ref{muSR}b. Both measurements show no temperature dependence of $\lambda$, implying the absence of local static electronic moments. The difference in the absolute values of $\lambda$ between the two measurements stems from the spin relaxation mechanism probed by each method \cite{muSRbook}. 

We have also performed longitudinal-field $\mu$SR measurements, as shown in Fig. \ref{muSR}c. At ZF we see a slow damping of the muon's polarization with time, that originates from frozen nuclear moments. Upon the introduction of a 50 G longitudinal field the signal completely flattens, suggesting that static local moments in our sample are much smaller than 50 G, which is a typical value for nuclear moments.
The $\mu$SR results are very clear, the ground state of 1T-TaS$_2$ is not an anti-ferromagnet.

\begin{figure}[ht]
	\centering
	\includegraphics[width=0.45\textwidth]{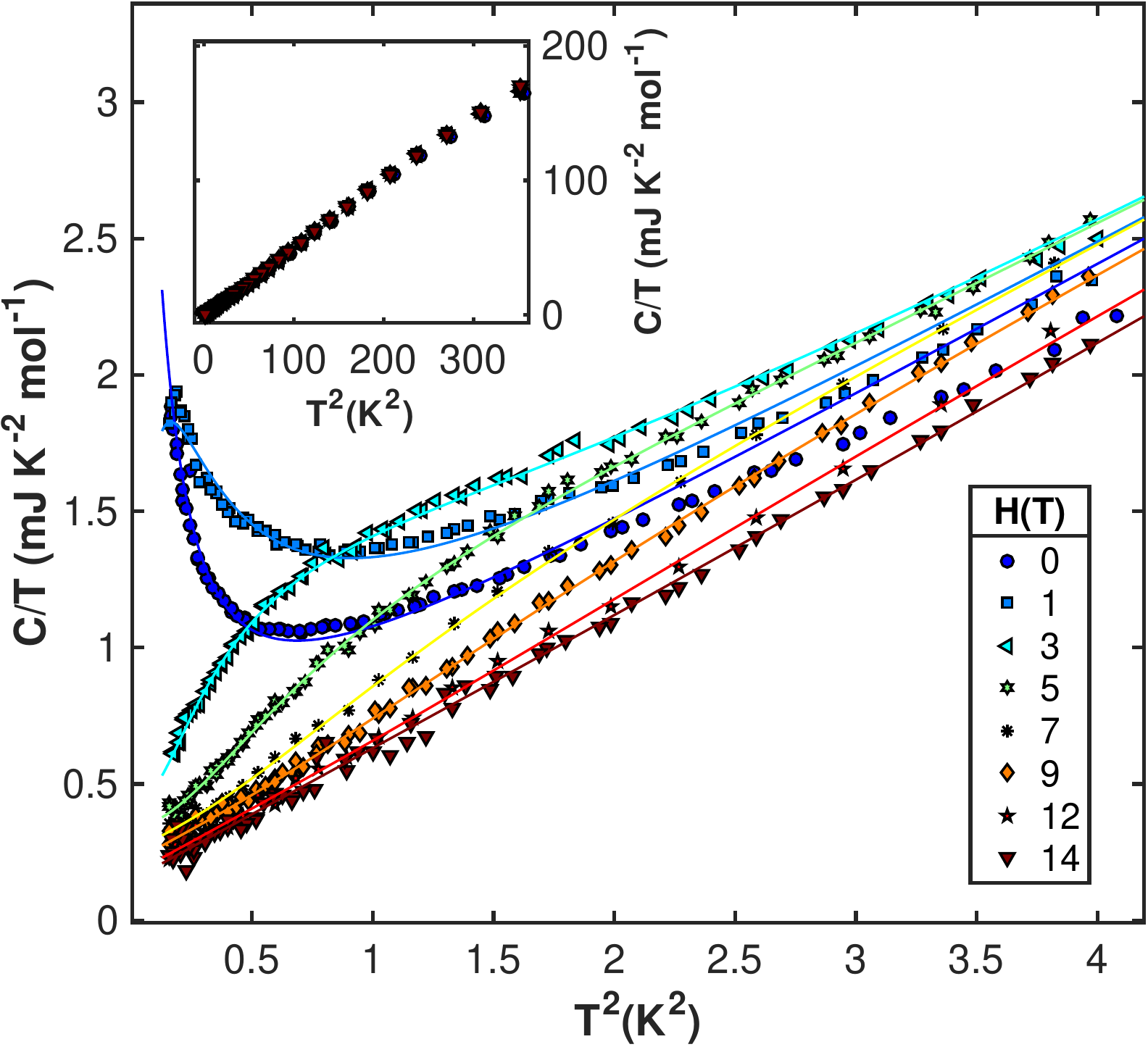}
	\caption{(Color online) Heat Capacity measurements of 1T-TaS$_2$ at eight magnetic fields, plotted as $C_P/T$ as a function of $T^2$. The  inset displays larger temperature range, which exhibit a linear relation according to the Debye form $C_P\propto T^3$. The main panel focuses on the low temperature regime where the deviation from the Debye behavior is noticeable. The solid lines are fits to equation (\ref{HC_model}) modelling the data using both a Schottky term and a linear contributions}
	\label{HC}
\end{figure}

Next we turn to our heat capacity data.
$C_P/T$ as a function of $T^2$ at various magnetic fields is presented in Fig. \ref{HC}. Above 2 K the main contribution to the heat capacity is from lattice vibrations as can be seen in the inset of Fig. \ref{HC}, yielding a Debye temperature of $\theta_D=156.3\pm 1.8K$, in accordance with former results \cite{Benda}. At lower temperatures and low magnetic fields the data exhibit a clear field-dependent deviation from the phonon contribution. Surprisingly, despite the insulating nature of the sample, a clear non-zero intercept of $C_P/T$ at $T=0$  is observed even for large magnetic fields. We therefore model the heat capacity as a sum of three contributions:
\begin{equation}
C_P=\gamma(H) T+\beta T^3+\frac{\alpha (H)}{T^2} \frac{\textrm{exp}(\frac{\tau(H)}{T})}{[\textrm{exp}(\frac{\tau(H)}{T})+1]^2}.
\label{HC_model}
\end{equation}
$\gamma(H)$ is a field dependent Sommerfeld term, usually attributed to fermionic quasi-particles. $\beta$ is the field independent Debye coefficient. The last term is the Schottky anomaly arising from a two energy-level system with a gap $\tau (H)$ using $\alpha(H)= n(H) k_B\tau^2(H)$ with $n$ the number of two level systems.
\par

The results of the fit to the zero-field data gives a two energy-level impurity density of $\simeq 1/2000$. This is comparable to the concentration of the nearly-free spins we extract from the Curie-like term in the susceptibility data assuming g$\sim$2.
The magnetic field affects the two level system gap as well as the number of such impurities (See supplementary for further details on $n(H)$ and $\tau(H)$).

\par
\begin{figure}[ht]
	\centering
 	\includegraphics[width=0.4\textwidth]{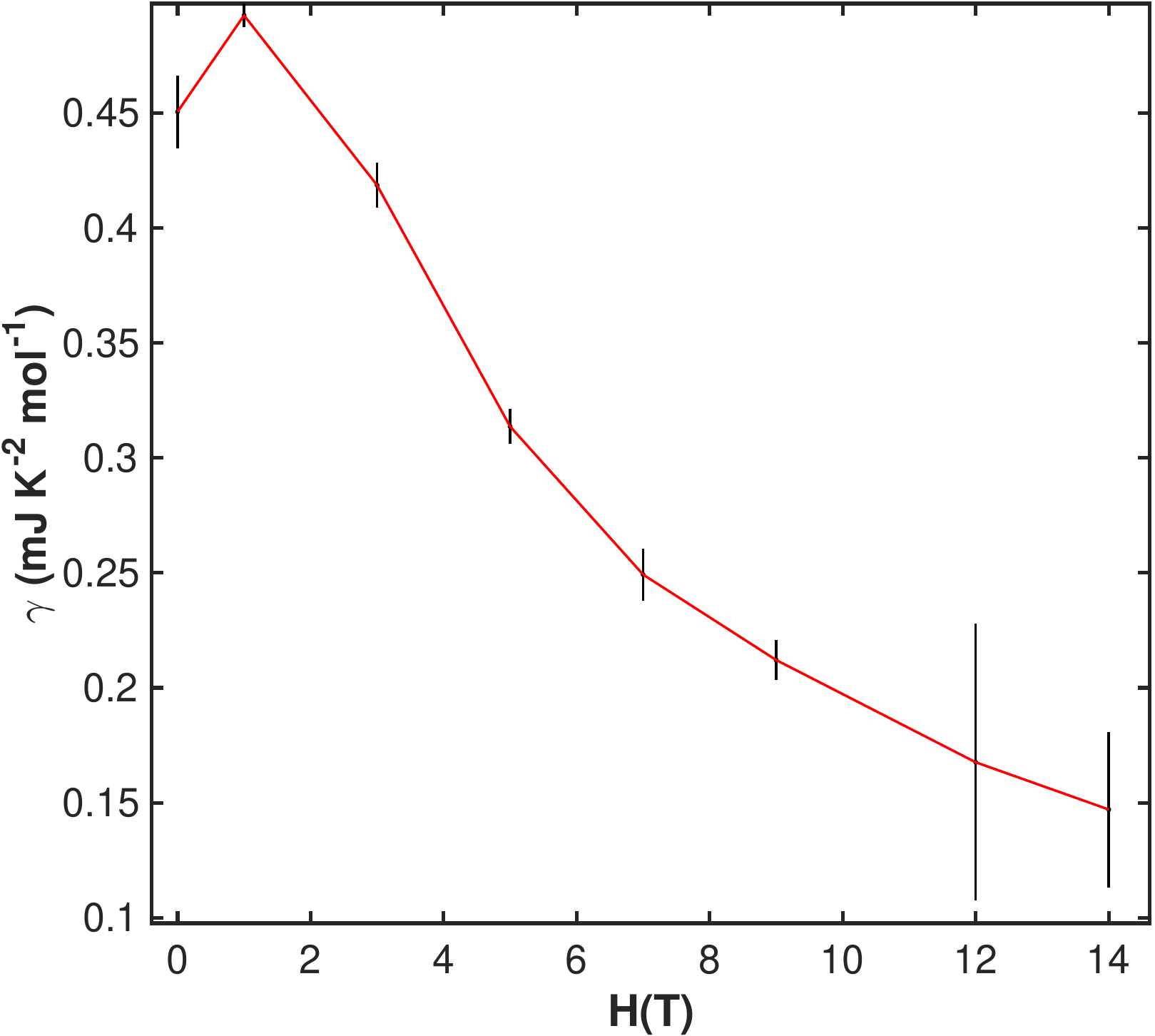}
 	\caption{(Color online) Field dependence of the heat-capacity linear term. $\gamma(H)$ decreases almost monotonically with the applied magnetic field.}
 	\label{GvsH}
\end{figure}

The main finding from the heat capacity measurements is the field-dependent linear term $\gamma(H)$, shown in Fig. \ref{GvsH}.
The origin of this term is unclear, but it is probably connected to the temperature-independent paramagnetic part of the spin susceptibility. Usually one would attribute the linear term to the contribution of free-carriers forming a Femi surface, but this system is highly insulating at low temperatures.
Assuming a 2D dispersion we extract from $\gamma$(0) an effective mass of $\simeq 6$ electron masses. Using a simplified tight-binding type model on a triangular lattice with a spacing of $\sqrt{13}a$ we estimate the bandwidth to be about 160meV.
On the other hand, the strong field dependence and the fact that $\gamma$ is reduced by a factor of 3 at 14T is in agreement with a much smaller bandwidth of the order of 0.1meV.

As with the temperature-independent paramagnetic term there are two ways to understand the linear term in the specific heat. The first is that this system is metallic along the c-axis,
as has been recently proposed \cite{Millis_PRB}.
However, the bandwidth we extract from the specific heat is smaller compared to the calculated 0.45eV  in \cite{Millis_PRB}.
Recent ARPES data \cite{Hofmann_ARPES} provide some support to c-axis conductivity scenario by showing some dispersion along the c-axis, although no crossing of the Fermi-level was found.
 The system is insulating with resistivity having an activation-type temperature dependence at low temperatures. If the insulating behaviour is a result of disorder than it is not clear if a linear term in the specific-heat is expected at all. There are systems which are insulating due to disorder and have a finite $\gamma$ term up to some level of disorder \cite{SiP_specificheat}. In addition, we do not expect a field dependence of $\gamma$ for a one dimensional conducting chain along the c-axis with such a bandwidth.
\par
The second possibility is that the origin of the linear term is from thermally excited spinons close to their Fermi-surface. In some models of QSL a gapless spectrum of excitations is expected \cite{Lee_AmpereanPair,Balents_QSL_review}. This is a result of a formation of band of spinons, whose width should be of the order of $J$ \cite{KBDET_NatPhys_2008,Lee}. A $J$ of the order of 100meV means that below the CCDW transition at about 200K the spins do not contribute to the Curie-like term, in agreement with the magnetic susceptibility results. A gapless QSL is expected to have a residual thermal conductivity $(\kappa_0)$ at T$\rightarrow$0K since the gapless spinons are itinerant excitations that carry entropy. A very recent paper claims that in 1T-TaS$_2$ $\kappa_0\sim0$ at very low temperatures \cite{TaS2_therm_cond}, raising doubts about the nature of the spin excitation. One possible explanation is that the gapless spin excitations are localized due to inhomogeneities and thus cannot conduct heat \cite{TaS2_therm_cond,arcon}. The strong field-dependence of $\gamma$ remains a puzzle in this scenario too.


To summarize, we studied the low temperature phase of 1T-TaS$_2$. We find an insulating behaviour both in-plane and out-of-plane, with a temperature dependence that seems to be activated-type. From the magnetic susceptibility we conclude that most of the sample (90-99.5\%) has a T-independent contribution, with a minority of the spins display a Curie-like behavior. By performing $\mu$SR measurements we have verified that there is no magnetic ordering down to 20mK. The main finding is the specific heat linear term which implies the existence of gapless excitations in 1T-TaS$_2$. These excitations have a bandwidth of $\sim$160meV. Finally, $\gamma$ changes significantly with an applied magnetic field. This field dependence remains unexplained.

We are indebted to P.A. Lee for helpful discussions and for a critical reading of the manuscript. We are grateful to E. Sela, M. Goldstein, B. Shapiro, D. Podolsky and N. Lindner for helpful discussions. This work was supported by the Israeli Science Foundation (Work at the Technion grant 320/17 and work at TAU grant 382/17).
Part of the $\mu$SR measurements were performed at the Swiss Muon Source (S$\mu$S), at the Paul Scherrer Institute in Villigen, Switzerland. Experiments at the ISIS Muon Source were supported by a beam-time allocation from the Science and Technology Facilities Council.

\bibliography{TaS2_QSL_bib}

\end{document}